\documentclass[sigconf, nonacm]{acmart}
\usepackage{graphicx}   
\usepackage{float}      
\usepackage{tabularx}   
\usepackage{booktabs}   
\usepackage{hyperref}
\AtBeginDocument{%
  }

\begin{document}

\title[Quantifying Qualitative Insights]{\huge Quantifying Qualitative Insights: Leveraging LLMs to Market Predict}

\author{Hoyoung Lee}
\affiliation{%
  \institution{Hankuk University of Foreign Studies}
  \country{Republic of Korea}
}
\email{greenday0021@hufs.ac.kr}

\author{Youngsoo Choi}
\affiliation{%
  \institution{Hankuk University of Foreign Studies}
  \country{Republic of Korea}
}
\email{choiys@hufs.ac.kr}

\author{Yuhee Kwon}
\affiliation{%
  \institution{Tech University of Korea}
  \country{Republic of Korea}
}
\email{yuheekwon@tukorea.ac.kr}

\renewcommand{\shortauthors}{Lee et al.}

\begin{abstract}
Recent advancements in Large Language Models (LLMs) have the potential to transform financial analytics by integrating numerical and textual data. However, challenges such as insufficient context when fusing multimodal information and the difficulty in measuring the utility of qualitative outputs, which LLMs generate as text, have limited their effectiveness in tasks such as financial forecasting. This study addresses these challenges by leveraging daily reports from securities firms to create high-quality contextual information. The reports are segmented into text-based key factors and combined with numerical data, such as price information, to form context sets. By dynamically updating few-shot examples based on the query time, the sets incorporate the latest information, forming a highly relevant set closely aligned with the query point. Additionally, a crafted prompt is designed to assign scores to the key factors, converting qualitative insights into quantitative results. The derived scores undergo a scaling process, transforming them into real-world values that are used for prediction. Our experiments demonstrate that LLMs outperform time-series models in market forecasting, though challenges such as imperfect reproducibility and limited explainability remain.
\end{abstract}

\ccsdesc[500]{Computing methodologies~Natural language processing}

\keywords{Stock Market Forecasting, Large Language Models, Explainable AI, Prompt Engineering}

\maketitle

\section{INTRODUCTION}
Financial markets are rich in both structured data, such as price movements, and unstructured data, like analyst reports and news articles. Ongoing innovations in LLMs have transformed the landscape of financial analytics \cite{zhao2024revolution}. By effectively utilizing contextual information, LLMs can make more accurate and sophisticated inferences \cite{brown2020llm}. This enhanced reasoning capability has led to impressive performance across various financial reasoning benchmarks \cite{li2023chatgpt}, as well as in rigorous assessments like the CFA exam \cite{callanan2023can}. Moreover, LLMs have been shown to outperform human analysts in predicting earnings changes \cite{kim2023llms}.
In addition to their strong reasoning abilities, LLMs are proficient at learning from and handling both numerical and textual data, having been trained on vast amounts of text from the internet across a wide range of topics, styles, and formats, enabling them to effectively process and integrate multi-modal data \cite{wu2023bloomberg}. This capability is particularly crucial in the financial domain, where even text-based data, such as financial reports and news articles, often contain important numerical information like prices and financial statements. Accurately analyzing these mixed data types is essential for making informed decisions, and LLMs’ ability to handle both numerical and textual data underscores their potential for tackling complex financial tasks.

However, efforts to harness LLMs' multi-modal capabilities by integrating numerical and textual data have faced challenges. Studies, such as those on using ChatGPT for zero-shot multimodal stock price prediction, reveal that these models often underperform compared to traditional methods like logistic regression, primarily due to a lack of sufficient context when fusing multimodal information \cite{xie2023wall}. While some studies provide extensive context to LLMs for financial predictions—such as incorporating historical stock data, company metadata, and financial news—these approaches often require substantial amounts of contextual information, which can be both computationally expensive and resource-intensive to obtain. Furthermore, overly complex contexts may hinder reasoning, making it difficult for the models to draw accurate inferences \cite{yu2023harnessing}. Additionally, due to the inherent characteristics of LLMs, challenges related to reproducibility persist, as the consistency of outputs across different inferences has not been deeply explored. This lack of thorough investigation into whether LLMs can consistently produce reliable results further complicates their application in accurate financial forecasting. 

To address these challenges, this study utilizes daily reports from securities firms to generate high-quality contextual information. These reports are segmented into text-based key factors and combined with numerical data, such as price information, to form comprehensive context sets. The few-shot examples are carefully constructed by considering temporal information, ensuring that they incorporate the most relevant data based on the query time. Furthermore, we have developed a systematic new prompt designed to assign scores to these key factors, converting qualitative insights into quantifiable data using a Likert scale scoring system. The resulting scores are then scaled to reflect real-world numerical values, making the qualitative analysis both measurable and actionable for prediction. 
Although LLMs are large and complex, often functioning as black-box models, we aim to enhance their explainability by guiding the generation process with rationale. This approach increases transparency in the reasoning process, making it easier to understand how the model arrives at its conclusions. By doing so, we improve interpretability and ensure the outputs are more comprehensible, while also striving for reliability in the model’s predictions.

\section{RELATED WORK}
\subsection{Traditional Methods}
Numerous researchers have used autoregressive integrated moving average (ARIMA)-type models to forecast prices and their returns \cite{box2015time, pokou2024hybrid}. Since these methods have limitations in capturing nonlinear relationships, to address this issue, researchers have applied deep learning techniques such as long-short term memory (LSTM) \cite{lim2021time}. However, there are still opportunities to improve prediction accuracy using qualitative variables such as financial reports.

\begin{figure*}[htbp]
    \centering
    \includegraphics[width=\textwidth]{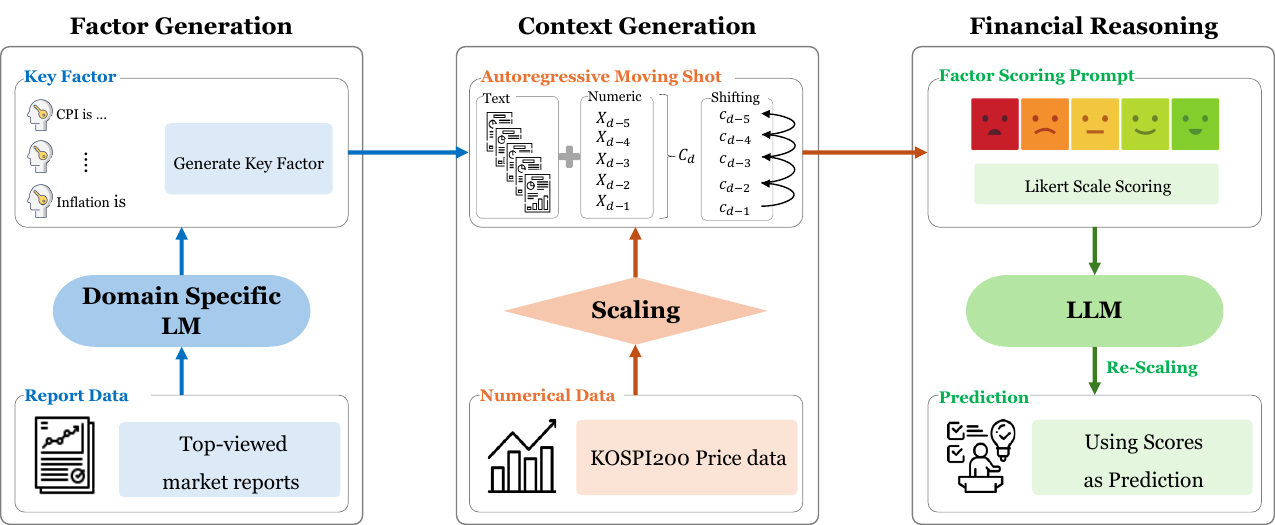} 
    \caption{The methodology involves three steps: generating key factors with a domain-specific model, combining them with price data to create autoregressive moving shots, and using factor scoring prompt for prediction. This process converts qualitative insights into quantitative results.}
    \label{fig:process}
\end{figure*}

\subsection{LLMs in Finance}
LLMs are revolutionizing the landscape of financial analysis with their diverse applications. Has been initiated Investigations into the financial reasoning capabilities of LLMs, providing evidence of their emerging potential to execute complex reasoning tasks within the financial domain \cite{son2023beyond}. This transformation is exemplified by automating complex financial workflows \cite{zeng2023flow}. Advanced prompting techniques are leveraged to extract and elucidate critical factors from financial news, greatly improving their readability and the explainability of influences on stock movements \cite{wang2024llmfactor}. The study showed how these models adeptly interpret the nuanced language of Korean financial reports to aid in strategic investment planning \cite{kim2023llms}. Despite these advancements, LLMs face significant challenges with hallucinations in output, prompting the need for continuous research and the application of innovative methods to ensure model reliability \cite{kang2023deficiency}. Additionally, LLMs improve the process of extracting information from the Q\&A portions of financial reports, reducing hallucinations and increasing both the accuracy and scalability of insights \cite{sarmah2023towards}. This suite of applications underscores the growing importance and utility of LLMs in refining and advancing financial services.
\subsection{Time-series forecasting with LLMs}
The sequences of numerical data as inputs for LLMs are directly used to make predictions \cite{gruver2024large}. Meanwhile, time series prediction through a prompt-based approach is explored in \cite{xue2023promptcast}, while additional efforts such as \cite{jin2024time} utilize paraphrasing trends in numerical data into prompts to enhance forecasting. For financial time series forecasting, \cite{yu2023harnessing} utilizes prompts to generate summaries and key phrases from LLMs, integrating various data sources to improve time-series predictions. A method to fine-tune LLMs using a self-reflective agent and Proximal Policy Optimization, which enables the autonomous generation of explainable stock predictions, is introduced in \cite{koa2024learning}. LLMs can predict stock market returns from news headlines more effectively than traditional forecasting methods, leveraging enhanced reasoning capabilities. It highlights that while basic models struggle with accurate forecasting, return forecasting emerges as a distinct capacity of larger LLMs \cite{lopez2023can}.
\subsection{Prompt Engineering}
Early work in this area includes the development of open-source models specifically trained on financial data, such as \cite{yang2023fingpt}, laying the groundwork for subsequent innovations. While there are models fine-tuned for the financial domain like \cite{xie2023pixiu}, general-purpose LLMs also demonstrate strong performance in reasoning when appropriately guided by prompt engineering \cite{li2023chatgpt}. Chain-of-Thought (CoT), a prompting method that guides models through step-by-step reasoning, enables LLMs to perform high-level reasoning tasks \cite{wei2022chain, kojima2022large}. Additionally, it has also enabled LLMs to perform quantitative reasoning \cite{lewkowycz2022solving}. However, these models may still experience hallucination issues \cite{kang2023deficiency}. To mitigate this, the use of retrieval-augmented generation \cite{lewis2020retrieval} maximizes the use of external knowledge, leveraging LLMs’ ability to incorporate and utilize such information through In-Context Learning (ICL)\cite{dong2022survey}.

\section{METHODOLOGY}
This section outlines how we enhance LLMs' financial reasoning and transition from qualitative to quantitative results, as detailed in Figure \ref{fig:process}, which illustrates the overall process.

\subsection{Data Collection}
Market information is collected to provide context for the financial reasoning of LLMs. Specifically, daily reports published by securities firm analysts, which are publicly available on Naver Finance\footnote{\url{https://finance.naver.com/research/} provides various financial data such as \url{https://finance.yahoo.com/}}, are utilized. These reports are commonly used by individual investors for research and investment purposes. The collected data, denoted as  $R_d$, where $d$ represents the date of the information, includes the top three most-viewed reports. Due to resource constraints and the need to manage context length effectively, the top three reports are selected based on view counts. This selection reflects individual investor preferences and ensures higher quality compared to other sources.

\subsection{Multiple Trials Generation}
In LLMs text generation, the model selects tokens one by one based on probabilities assigned to each possible next word. Greedy decoding is a common approach where, at each step, the model chooses the most probable token. While this method produces deterministic results, it often fails to explore alternative generation paths, leading to sub-optimal performance \cite{wang2024chain}.   

Increasing the temperature in the sampling process is necessary to expand the generation paths, as it allows the model to explore a wider range of possible outputs by giving less probable tokens a higher chance of being selected. Since the generated results are used for prediction, a low temperature was adopted instead of a high one to introduce minimal variation without causing significant randomness. Using the same information and prompts, \(k\) trials were conducted, and the median result among them was selected. This approach increases the diversity of inference paths while maintaining the consistency needed for reliable predictions, and the results examining differences across the trials can be found in Section \ref{sec:trials}.

\subsection{Factor Generation}
Using raw report data is inefficient due to the volume and complexity, making it difficult to trace the reasoning process. To address this, the information is segmented by identifying key factors that influence future market fluctuations, and concise summaries are generated for each factor. This process is carried out using a domain-specific model, Finance-Llama-8B \cite{cheng2023adapting}.  Since the generation of key factors must minimize information loss, a domain-specific model is employed to ensure that the task is tailored to the specific domain \cite{yang2023empower}. This factor-based approach outperforms sentiment-based methods in text utilization \cite{wang2024llmfactor}.
\begin{equation}
\mathcal{D} \left( \text{combination of reports} \{ R_{d} \} \right) = F_d^k = \left\{ f_{d, 1}^k, f_{d, 2}^k, \dots, f_{d, 10}^k \right\}
\end{equation}
where $\mathcal{D}$ denotes the domain-specific model. Each $F_d^k$  represents a set of 10 key factors extracted from the combined reports. The segmented key factor generation process facilitates financial reasoning by allowing focused analysis on each individual factor. In Section 5.2, the results of scoring each factor are presented, with transparent reasoning provided for the assigned scores, ensuring accountability and clarity in the decision-making process.

\subsection{Autoregressive Moving Shot}
To utilize the ICL capability of LLMs, past data is structured as few-shot examples to form the context. Key factors alone often miss critical details, such as price trends. Therefore, numerical data, like price changes, is combined with the key factors to provide a more complete context. Additionally, to ensure the model makes predictions using the given context, a time lag is introduced between the key factor information and the actual price changes. By presenting the price changes from a future point relative to the key factor within the same context, better predictive performance is facilitated.
\begin{equation}
X_d(l)=g\left(S_{d+1}-S_{d-l}\right)
\end{equation}

Here, \(S_{d}\) represents the price at time \(d\). Price changes are denoted as \(X_d(l)\), representing the difference between future and past prices over a given look-back window. The look-back window is denoted by \(l\). To account for various past values, the look-back window is utilized. Detailed experimental settings are described in Section 4.5. The function \(g\)  is applied to truncate the decimal points of the price difference, reducing the complexity of the numbers for easier processing.
In constructing the context, key factors and corresponding price changes for the given date are bundled into a single set, denoted as the context set \(C_d^k\). This context set serves as the foundation for modeling the predictive relationship between key factors and price trends.
\begin{equation}
C_d^k = C_d^k(l) = \left( F_d^k,  X_d(l) \right)
\end{equation}

The \textit{Autoregressive Moving Shot} approach is designed to continuously shift the context as new data becomes available. To maintain a comprehensive and temporally relevant context, the past 5 days’ context sets are used, reflecting weekly trends common in financial data. This is expressed as:
\begin{align}
\text{5-Shot}_d=\left\{ C_{d-1}^k, C_{d-2}^k, \dots, C_{d-5}^k \right\}
\end{align}

As the prediction date updates, the context sets move autoregressively. This means that each day, a new example is incorporated while older ones are removed. Using a 5-Shot captures the short-term weekly trends, which are crucial in financial analysis. This method ensures that the most recent and relevant information is always included, enhancing the predictive performance of the model. The autoregressive nature of this approach allows the model to dynamically adjust to changing data patterns, ensuring the context remains aligned with current data. This strengthens temporal alignment and provides a continuously updated and relevant predictive framework.
\begin{figure}[htbp]
    \centering
    \includegraphics[width=\linewidth, height=0.47\textheight]{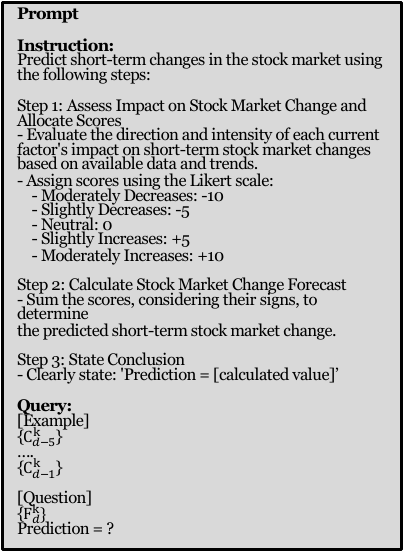} 
\caption{Factor Scoring Prompt is presented with 5-Shot.}    \label{fig:prompt}
\vspace{-3mm} 
\end{figure}
\subsection{Factor Scoring Prompt}
To convert qualitative information into quantitative results, we employ a modified Likert scale specifically designed for financial analysis. While traditional Likert scales measure sentiments on a linear spectrum, our framework categorizes responses into five segments: Moderately Decreases, Slightly Decreases, Neutral, Slightly Increases, and Moderately Increases.
Assigning numerical values to the inferences drawn from each key factor follows these five segments as the criteria for scoring. The assigned scores are later scaled to align with actual price fluctuation data, ensuring a more accurate reflection of market movements. This translation of text-based insights into measurable data ensures that the finding is practically applicable.

\textit{Factor Scoring Prompt} provides step-by-step guidelines to draw out the LLM’s reasoning capabilities. Question is structured alongside \textit{Autoregressive Moving Shots}, ensuring that the model processes both historical and current context effectively. To reduce the complexity of numerical operations, integers are used instead of decimals when assigning scores. Detailed examples of the prompts can be found in Figure \ref{fig:prompt}.

\subsection{Value-to-Score Scaling and Re-Scaling}
The scoring framework assigns integer values to various factors based on qualitative assessments derived from \textit{Factor Scoring Prompt}. Each factor is evaluated according to specific criteria, and these scores are then summed to calculate the total score for the day. In order to make the scores comparable with actual financial data, a scaling process is necessary. 
First, the range of the total score is defined. Let \(\text{score}_{d,f}\) represent the score for factor \(f\) on day \(d\). The total score for all factors on day \(d\), denoted as \(\operatorname{TotalScore}_d\), is the sum of individual factor scores:
\begin{equation}
\operatorname{TotalScore}_d = \sum_{f=1}^{N_f} \text{score}_{d,f}
\end{equation}
The total score is constrained by the following range:
\begin{equation}
-\mathrm{score}^{\max} \times N_f \leq \operatorname{TotalScore}_d \leq \mathrm{score}^{\max} \times N_f
\end{equation}

Here, \(\mathrm{score}^{\max}\) is the maximum possible score for each factor, and \(N_f\) is the total number of factors. To align the total score with real-world financial data, a scaling method is applied. The minimum and maximum values of \(X_d\), $X_d^{(\min)} \,\&\, X_d^{(\max)}$, are calculated over a rolling 21-day window\footnote{A period of 21 days is suitable for capturing short-term patterns and trends, and it represents one month in the financial markets.} to account for potential upper and lower bounds of future price movements based on recent short-term prices. These bounds help manage outliers that may arise when assigning numerical scores with the LLMs, ensuring that extreme values do not disproportionately affect the final score. To handle outliers, a distribution scaling multiplier \(m_{\text{dist}}\)\footnote{Under the assumption of a t-distribution with thicker tails than a normal distribution, the occurrence frequency was extended from 21 days to 100 days.} is introduced, which is computed as:
\begin{equation}
m_{\text{dist}} = \frac{\mathrm{qt}(0.99, 21)}{\mathrm{qt}(0.95, 21)} = 1.463
\end{equation}
Using this multiplier, a factor adjustment multiplier \(m_{\text{f,d}}\) is computed to scale the score based on the observed data:
\begin{equation}
m_{f,d-j} = \min\Bigl(\bigl| \tfrac{-\mathrm{score}^{\max} \cdot N_f}{m_{\text{dist}} \cdot X_d^{(\min)}} \bigr|,\,\, \bigl| \tfrac{\mathrm{score}^{\max} \cdot N_f}{m_{\text{dist}} \cdot X_d^{(\max)}} \bigr| \Bigr)
\end{equation} 
This factor adjustment multiplier is applied to the price changes in the 5-shot, denoted by $\{X_{d-j} \}_{j=1}^5$, and replaced by scaled values calculated using the following equation:
\begin{equation}
X_{d-j}^{\text{scaled}} = X_{d-j} \cdot m_{f,d-j}, \quad j = 1, \dots, 5
\end{equation}

The scaled values are then used as input for LLMs to predict market movements.
Finally, to align the predicted score with the original data scale, a rescaling process is performed. The total score generated by the LLM is adjusted by the factor adjustment multiplier \(m_{\text{f,d}}\), ensuring that the scaled score is mapped back to the real-world financial context. This rescaled value is computed as follows:
\begin{equation}
X_d^{(\text{re-scale})} = \frac{\operatorname{TotalScore}_d}{m_{f,d}}
\end{equation}
By applying this rescaling, the adjusted score can be effectively utilized to represent price changes.

\section{EXPERIMENTS}

\subsection{Experimental Period}
The data used in our experiments covers the period from June 1, 2023, to May 31, 2024, with daily-level collection and analysis. Due to the computational demands of running multiple inferences, we are constrained by practical limitations in extending the experimental period further.

\subsection{Baselines}
We compare our method against two commonly used time series models:

\paragraph{ARIMA} is a statistical model that combines autoregression, differencing, and moving averages for time series forecasting.

\paragraph{LSTM} is a deep learning model designed to capture long-term dependencies in sequential data using memory cells.

\subsection{Large Language Models}
Two LLMs are evaluated to test their effectiveness in financial reasoning:

\paragraph{GPT-4-Turbo} \cite{achiam2023gpt}, an improved version of GPT-4 developed by OpenAI, enhances performance and efficiency.

\paragraph{LLaMA3} (MetaLlama-3-70B-Instruct) \cite{touvron2023llama}, an open-source model from Meta, is designed for accurate instruction following in various tasks.

\subsection{Performance Evaluation}
The experiments utilize the KOSPI200 index, which represents the daily closing prices of the top 200 companies listed on the Korean Exchange. The task involves predicting KOSPI200 price changes, and performance is evaluated using the following three metrics: Accuracy (ACC), Matthews Correlation Coefficient (MCC), and Root Mean Squared Error (RMSE), following prior work \cite{yu2023harnessing}.

\subsection{Look-back Window}
This experiment considers both the forward-looking and retrospective nature of reports. Reports are typically written after specific events, inherently reflecting past events and introducing a time lag in the availability of information. Given the prompt structure that includes 5-shot context information, it is crucial to note that the 5-shot cannot include information beyond the question date. Therefore, we fix the future point one day ahead and extend the look-back windows $l \in\{1, 2, 3\}$ while predicting $\hat{X}_{d+1}(l)$. This approach allows us to assess how well textual insights derived from extended historical contexts can explain subsequent price movements, considering both the forward-looking and retrospective aspects of financial reports. The predicted values \(\hat{X}_{d+1}(l)\) are then compared with the actual values \(X_{d+1}(l)\) to evaluate the models. By progressively extending the look-back window, we can examine whether reasoning effectiveness emerges.

\section{RESULTS}
\subsection{Multiple Trials Result} \label{sec:trials}
By setting  \(k = 5\) and a temperature of 0.2, we examine whether consistency exists between trials. Each trial uses identical information and prompts to ensure a fair comparison. To assess the similarity of the factors  \(F_d^k\)  generated in each trial, we embed the text outputs and compute pairwise cosine similarities. Table \ref{emb_table} presents the mean and standard deviation of these similarities using both TF-IDF and semantic embedding\cite{lee2024nv}. The high cosine similarity scores indicate that the factors generated across different trials are highly similar, confirming consistency in the model’s outputs despite the introduced randomness.
\begin{table}[h] 
\centering
\begin{tabular}{@{}c|cc@{}}
\toprule
Method             & Mean & Std  \\ \midrule
TF-IDF Embedding   & 0.82 & 0.09 \\
Semantic Embedding & 0.92 & 0.04 \\ \bottomrule
\end{tabular}
\caption{Pairwise cosine similarities computed via embeddings, showing consistency across trials.}
\label{emb_table}
\vspace{-10pt}
\end{table}

\begin{figure}[H]
    \centering
    \includegraphics[scale=0.8]{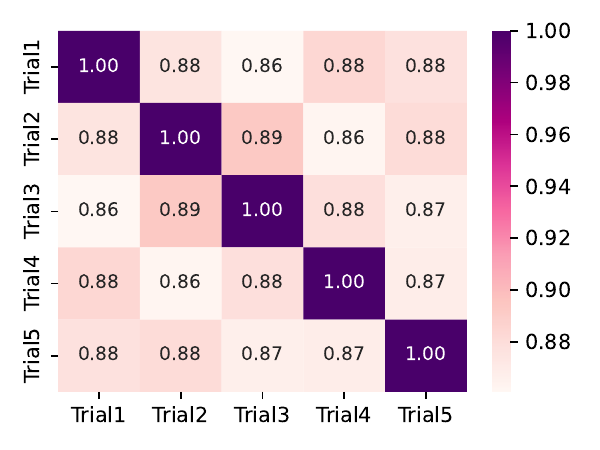} 
    \caption{Heatmap of pairwise correlation coefficients between trials, showing strong consistency with coefficients over 0.85}
    \label{heatmap}
\end{figure}

We further analyze the total score correlations between trials over the entire experimental period. Figure \ref{heatmap} presents the pairwise correlation coefficients among the five trials. Correlation coefficients exceeding 0.85 demonstrate strong consistency between trial results. This shows that increasing the temperature to 0.2 expands the LLM’s generation paths while introducing minimal randomness. By adopting a low temperature, we ensure the generated results remain consistent yet allow for slight variation. These findings support our approach of using multiple trials with a low temperature. The model maintains reliable performance, and the slight randomness enhances diversity without compromising prediction stability, validating our strategy of balancing exploration and predictability.

\subsection{Financial Reasoning Result}
\begin{figure}[htbp]
    \centering
    \includegraphics[width=1\linewidth, height=0.6\textheight]{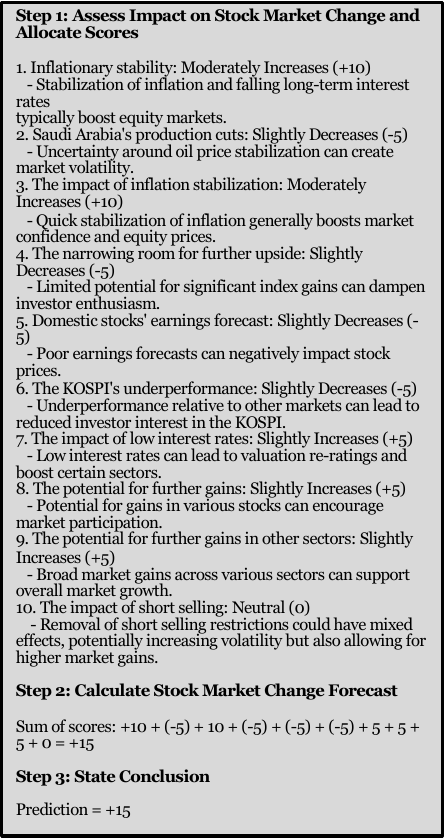} 
    \caption{LLM-generated text with Factor Scoring Prompt, along with rationale for transparency.}
    \label{output}
\vspace{-10pt}
\end{figure}

\begin{table*}[t]
\centering
\renewcommand{\arraystretch}{1.2} 
\begin{tabular*}{\linewidth}{@{\extracolsep{\fill}}cccccccccc@{}}
\toprule
            & \multicolumn{3}{c}{Look-Back ($l=0$)} & \multicolumn{3}{c}{Look-Back ($l=1$)} & \multicolumn{3}{c}{Look-Back ($l=2$)} \\ \cmidrule(lr){2-4} \cmidrule(lr){5-7} \cmidrule(lr){8-10} 
            & ACC     & MCC      & RMSE   & ACC     & MCC      & RMSE   & ACC     & MCC     & RMSE    \\ \midrule
ARIMA       & 0.53    & 0.00     & 3.65   & 0.42    & -0.17    & 8.98   & 0.64    & 0.27    & 7.04    \\
LSTM        & 0.47    & -0.06    & 3.82   & 0.42    & -0.15    & 6.42   & 0.62    & 0.24    & 6.48    \\
LLaMA3      & 0.49    & 0        & 5.71   & 0.66    & 0.34     & 6.3    & 0.66    & 0.33    & 6.74    \\
GPT-4-Turbo & 0.49    & -0.01    & 5.76   & 0.68    & 0.36     & 6.68   & 0.69    & 0.38    & 6.65    \\ \bottomrule
\end{tabular*}
\caption{Performance comparison of LLMs and baseline across different Look-Back periods. As $l$ increases, more past period is included in the prediction period. LLMs outperform the baseline models at Look-Back $(l=1)$, but from Look-Back $(l=2)$ onward, the past data becomes too diluted, and the meaning of the prediction becomes unclear.}
\label{eval_table}
\end{table*}
The financial reasoning results are examined in detail through the analysis of a single sample. The LLMs utilize the provided context and prompts to generate explainable inferences for each factor, as illustrated in Figure \ref{output}. The information from the reports is segmented into key factors, and a structured scoring system is applied to each factor, facilitating a comprehensive evaluation. Factor scoring follows the Likert scale method, as outlined in the Factor Scoring Prompt, enabling a systematic assessment process. Each score is accompanied by a rationale that offers detailed explanations aligned with the respective scores. This alignment ensures consistency and enhances the transparency of the decision-making process. By breaking down report information into distinct factors and assigning scores, the approach enables a more holistic and well-rounded judgment. The rationale not only justifies the scores but also provides valuable insights crucial for understanding the financial implications of each factor. This method effectively quantifies qualitative information, significantly improving the practical utility of these insights for financial analysis and decision-making. Through the combination of scoring and rationale, the process supports a more informed and reliable evaluation, aiding stakeholders in making data-driven decisions.

\subsection{Performance Evaluation Result}
The reasoning effectiveness of LLMs is evaluated over time through a multi-look-back analysis, focusing on short-term predictions. Metrics such as ACC and MCC are used to evaluate the direction of market movements, which are crucial for identifying optimal buy and sell points. As shown in Table \ref{eval_table}, in the one-day ahead prediction Look-Back $(l=0)$, all models, including LLMs, produce relatively insignificant results. This outcome is expected, as the textual data provided to LLMs is based on reports written after specific events occur. These reports are not suitable for immediate predictions and require consideration of the time lag in the availability of information.

As the Look-Back increases, the prediction period expands to include both past and future data, allowing for a more comprehensive analysis of the changes leading up to the future event. In Look-Back $(l=1)$, the window extends to cover a two-day period, from one day ahead to one day prior, to account for the time lag in the availability of information. Unlike time series models, which show a decline in performance as the period increases, LLMs demonstrate improvement. This suggests that LLMs effectively capture the transitions from past to future, leveraging both the forward-looking nature of financial reports and the retrospective timing of information acquisition. In addition to improved accuracy metrics, it is also observed that the RMSE for Look-Back $(l=1)$ is lower for LLMs compared to time series models.

Once the Look-Back extends beyond $(l=1)$, increasing the past period dilutes the proportion of relevant past information in the prediction period, shifting the focus further away from the essence of predicting future outcomes. Additionally, the accumulation of past data reduces the standard deviation of the target variable, giving time series models an advantage by simplifying the complexity of predictions. At $(l=2)$, the performance difference between time series models and LLMs becomes minimal, indicating that both methods capture similar predictive value within this period. Therefore, no further extension of the Look-Back is applied.
\section{CONCLUSION}
The integration of LLMs into financial analytics represents a major advancement in managing complex reasoning and decision-making. This study demonstrates the potential of LLMs to derive valuable insights by combining numerical and textual data. By applying financial domain knowledge during context crafting, ICL and CoT techniques, LLMs are able to deliver more accurate and actionable predictions tailored to the financial sector. The experimental results further highlight that LLMs consistently outperformed traditional time series models, particularly in capturing nuanced market trends and generating more reliable forecasts. This superior predictive performance demonstrates the potential of LLMs to be effectively utilized as decision-making tools in financial analytics. Additionally, generating predictions along with rationale provides greater transparency, enabling a clear understanding of how decisions and predictions are made.

However, challenges persist, particularly in achieving full reproducibility. While LLMs produce reasonably consistent results across trials, they do not always generate identical outcomes, highlighting limitations in achieving complete reproducibility. Moreover, while the study successfully transforms qualitative insights into quantitative measures through factor-based scoring, a limitation remains in how the justification for these scores is provided. Compared to model-based explainable methods, such as using model weights to justify scores, the approach of providing text-based rationales offers limited explainability.
\section{FUTURE WORK}
Future work aims to improve the explainability of LLMs in financial analytics by leveraging token-level probabilities during generation. While the model currently provides text-based rationales, it lacks transparency in how these are formed. By integrating token probabilities into the explanation process, predictions can be more clearly linked to quantitative metrics, offering a stronger, data-driven justification. This approach will enhance explainability.


\bibliographystyle{plain}
\bibliography{ref}

\end{document}